**Excitonic Complexes and Optical Gain in Two-Dimensional Molybdenum Ditelluride Well below Mott Transition**

Zhen Wang[1,2†], Hao Sun[1,2†], Qiyao Zhang[1,2], Jiabin Feng[1,2], Jianxing Zhang[1,2], Yongzhuo Li[1,2], and Cun-Zheng Ning[1,2,3*]

[1]Department of Electronic Engineering, Tsinghua University, Beijing 100084, China

[2]International Center for Nano-Optoelectronics, Tsinghua University, Beijing 100084, China

[3]School of Electrical, Computer, and Energy Engineering, Arizona State University, Tempe, AZ 85287, USA

*Correspondence to: cning@tsinghua.edu.cn; cning@asu.edu

†These authors contributed equally to this work.

**Abstract**

*Strong Coulomb interaction in 2D materials provides unprecedented opportunities for studying many key issues of condensed matter physics, such as co-existence and mutual conversions of excitonic complexes, fundamental optical processes associated with their conversions, and their roles in the celebrated Mott transition. Recent lasing demonstrations in 2D materials raise important questions about the existence and origin of optical gain and possible roles of excitonic complexes. While lasing occurred at extremely low densities dominated by various excitonic complexes, optical gain was observed in the only experiment at densities several orders of magnitude higher, exceeding the Mott density. Here, we report a new gain mechanism involving charged excitons or trions well below the Mott density in 2D molybdenum ditelluride. Our combined experimental and modeling study not only reveals the complex interplays of excitonic complexes well below the Mott transition, but also provides foundation for lasing at extremely low excitation levels, important for future energy efficient photonic devices.*

Dynamical processes of quasiparticles such as excitons and their various associated complexes (including charged excitons or trions, bi-excitons, and other highly correlated objects *(1)*) in solids are at the very core of fundamental condensed matter physics. The evolution of physical states from low to high carrier density involves *(2-5)* the insulating exciton gas, Bose-Einstein condensate (BEC) (2*),* co-existence of various excitonic complexes, crossover or transitions to conducting electron-hole plasmas (EHP), or electron-hole liquid (EHL) *(3)* through the Mott transition (MT) *(4)*. There remain many fundamental issues to be understood associated with the evolution of such quantum quasiparticles and the related phase transitions, especially in the intermediate density regime involving highly correlated complexes. On the other hand, mutual conversions of these excitonic complexes and related phase transitions are also profoundly linked to the natures of different optical processes in semiconductors as carrier density increases *(5,6)*. Light-semiconductor interaction thus plays important dual roles: As an information reporter, the interaction reveals the intrinsic dynamics of evolution, co-existence, and mutual conversion of various excitonic complexes in their passage towards degenerate EHP or EHL through MT, providing deep understandings of physical processes when carrier density is successively increased. At the same time, different conversion processes among various excitonic complexes provide novel absorption or emission mechanisms, forming the physical foundations for photonic functionalities including lasers, solar cells, light emitting diodes, and many other devices. While optical processes are well-understood in the two extremes: pure exciton gas and highly degenerate EHP, the intermediate stages involving various excitonic complexes are much less understood. Recent study *(7)* shows an interesting relationship between the Mott density (MD) and the transparency density, at which optical gain or stimulated emission occurs, especially in low dimensional systems. Limited study of the stimulated emission through excitonic complexes in the intermediate density regimes has produced both scientific discoveries and technological breakthroughs in semiconductor photonics research, such as optical gain based on excitons *(8,9)*, bi-excitons *(10,11)*, trions *(12,13)*, and polariton-scatterings *(14)* in III-V or II-VI compound semiconductors, or multi-exciton *(15)* and single-exciton *(16)* gain observed in nanocrystals. These relatively rare gain mechanisms have resulted lasing demonstrations at much lower pumping thresholds than conventional lasers which require higher pumping above MD.

The advent of 2D materials provides new impetus to the study of the issues discussed above, due to the dramatically weakened screening of Coulomb interaction. This allows the rich physics related to excitonic complexes, associated mutual conversions, optical transitions, and phase transitions such as MT, and device applications to be explored in a larger range of energies and carrier densities, at higher temperatures, with larger range of controlled parameters, and with more excitonic species to co-exist and mutually convert, than before in conventional III-V or II-VI semiconductors. Among many excitonic complexes are excitons, bi-excitons, and trions, an important type of quasiparticles consisting of an exciton and an electron or a hole. While the existence and basic optical features of these excitonic complexes have been investigated *(17-23),* the possibility of associated optical gain has not been studied. In related developments, couplings of 2D materials with photonic nano- and micro-cavities have attracted great interests, leading to demonstrations of 2D materials-based stimulated emission and lasing *(24-29)*. One of the most fundamental questions is the existence and origin or physical mechanism of optical gain, a prerequisite for lasing, in a 2D material. The lasing in these experiments occurs at very low pumping levels of a few W/cm$^2$ *(24,25),* or MW/cm$^2$ *(26)*. The corresponding carrier density levels are around $10^6$ or $10^{11}$ cm$^{-2}$. But the possibility of optical gain associated with various excitonic complexes below the MT density has not been explored. In several interesting recent developments, optical gain, MT, and the related giant bandgap renormalization were demonstrated at extremely high-density levels on the order of $10^{14}$ cm$^{-2}$ experimentally *(30)*, or theoretically *(31,32)*. The orders of magnitude discrepancy in carrier densities between the laser demonstrations and existence of optical gain after MT raises important questions. For example, does optical gain in 2D materials require to be in the degenerate EHP state or can it occur before MT, supported by excitonic complexes? More importantly, a fundamental question is if there are more general and profound interrelationships between various excitonic complexes and stimulated emissions among them in the intermediate density regimes. The answers to these questions provide not only key insights into the co-existence and mutual conversions of various excitonic complexes, but also physical basis for operating 2D materials-based photonic devices such as lasers at much lower excitation levels and much higher temperatures than in conventional semiconductor lasers.

In this paper, we study the co-existence and mutual conversions between excitons and trions in electrically gated mono- and bi-layer molybdenum ditelluride ($MoTe_2$) (see *Methods* section for details). Such electrical-gate control was previously used to study trions in a $MoS_2$ field-effect transistor *(17),* to explore the conversion dynamics between excitons and trions *(18-20)*, or to study light emitting diodes *(33,34)*. We conducted systematic micro-photoluminescence (μ-PL) and reflectance spectroscopy (see *Methods* section for details) on the electrically gated $MoTe_2$ devices using continuous wave (CW) lasers. Our intention was to probe the inner mechanisms of co-existence or transition kinetics among various excitonic complexes as pumping increases and to investigate the possibility, the origin, and physical mechanism of optical gain in such low-density regimes.

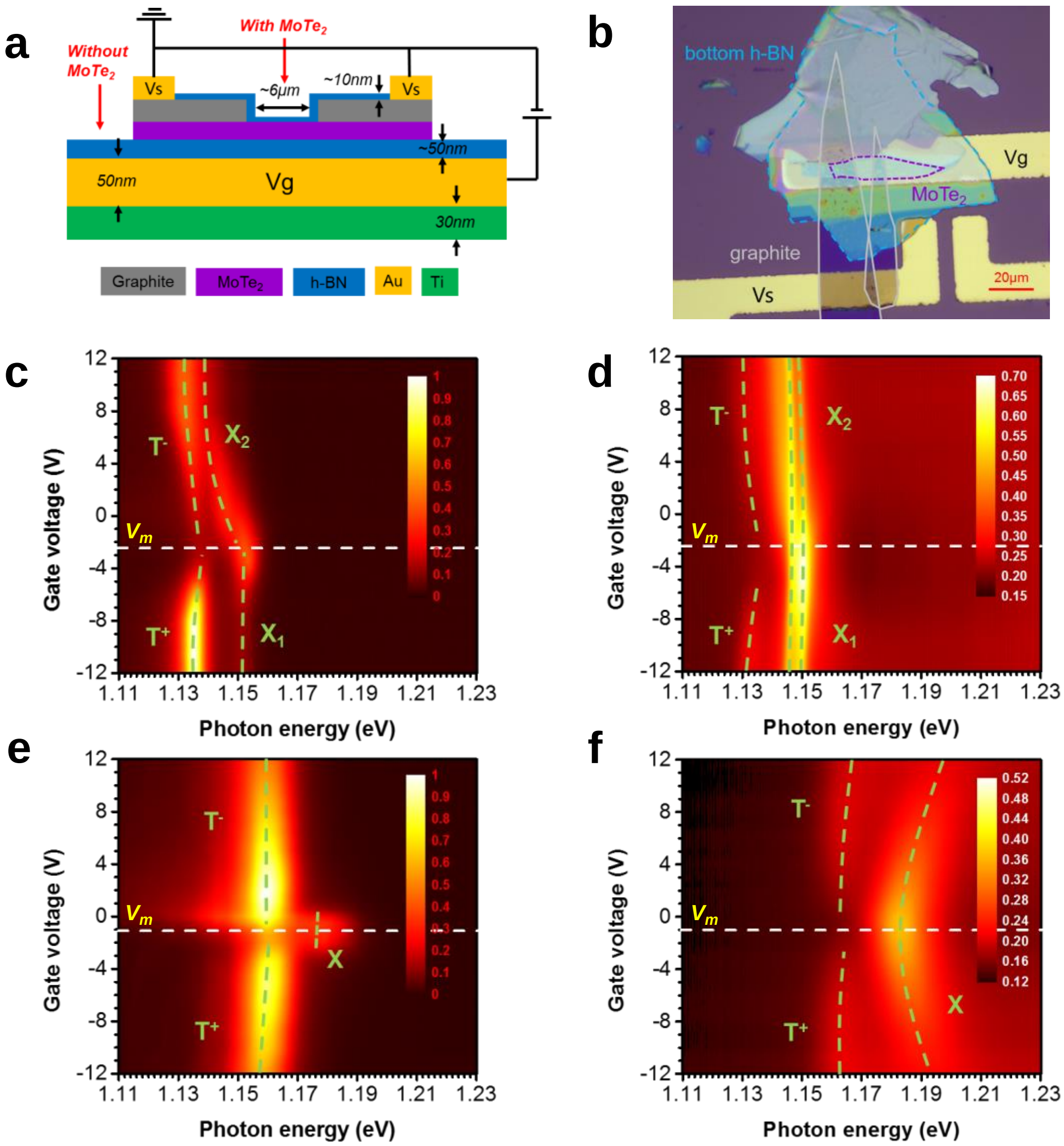

**Figure 1 | Sample structure and basic optical properties**. **a**, Schematic of the electrically-gated $MoTe_2$ device. The mono- or bi-layer $MoTe_2$ flakes of sizes typically ~10 µm are encapsulated with h-BN. The entire structure was placed on top of a $SiO_2$ (300 nm)/Si substrate covered with Au/Ti electrode. **b.** Optical microscope image of a fabricated device with $MoTe_2$ marked by the purple dashed lines. The grey and blue dashed lines indicate the regions of graphite contacts and h-BN, respectively. PL (**c** & **e**) and absorption (**d** & **f**) map in the plane of gate voltage and photon energy for bilayer (**c** & **d**) and monolayer (**e** & **f**) $MoTe_2$, respectively. Different excitonic species are well resolved, including excitons ($X_1$ & $X_2$), electron-trion ($T^-$) and hole-trion ($T^+$) states. The green dashed lines indicate the maxima of the spectral features, while the horizontal white dashed lines indicate the maxima of excitons and minima of trions.

Figure 1 shows a schematic diagram (a) and optical microscope image (b) of the electrically-gated sample structure with a bottom Au/Ti electrode (*Vg*), separated by a 50 nm h-BN film from the 2D $MoTe_2$ layer for electrical control of charges. The top contact (*Vs*) consists of two stripes of graphite separated by 6 µm to allow access of laser excitation and collection of reflectance from the $MoTe_2$. Absorption ( $\alpha(e)$ ) and optical gain spectra ( $g(e)$ ) at a given pumping level, $e$ , are related to differential reflectance (see Supplementary Materials, SM, Section S1 for more discussions) as follows:

$$g(e) = -\alpha(e) \propto \frac{\left[R(e,p,s) - R(0,p,0)\right] - R(e,0,s)}{R(0,p,0)} \tag{1}$$

where $R(e,p,s)$ represents reflection spectrum in the presence of CW-laser excitation ("$e$"), broad band probe ("$p$"), and the sample ("$s$"). A zero ("$0$") in the respective positions represents the absence of excitation, probe, or sample, as indicated in Fig. 1(a). It is important that PL ($R(e,0,s)$) is subtracted in the numerator of *equation (1)*, since, by definition, optical gain or absorption is the response of a material to a weak probe. This subtraction is important in CW experiments, as opposed to ultrafast pump-probe measurements, where PL process has not occurred during the short delay between the excitation and probe pulses. Thus, the intrinsic absorption ($\alpha(0)$) of a material without laser excitation is given by:

$$-\alpha(0) \propto \frac{\left[R(0,p,s) - R(0,p,0)\right] - R(0,0,s)}{R(0,p,0)} = \frac{R(0,p,s) - R(0,p,0)}{R(0,p,0)} \tag{2}$$

Equation (1) and (2) are referred to as differential reflectance and they are commonly used for the determination of optical gain or absorption of a thin-film. More detailed discussions about our specific structure are given in SM Section S2. The PL and absorption (processed using *equation (2)*) spectra are shown in Fig. 1 (c)-(f) for bilayer measured at 4 K (c & d) and monolayer $MoTe_2$ samples measured at 10 K (e & f), respectively, as gate voltage is varied between +/- 12V. Stacked line plots of the same data are reproduced in SM Section S3 (Fig. S 5), to more clearly display the spectra features and mutual conversions of excitonic species. We see from both PL and absorption spectra that the intensity and spectral positions of trions and excitons can be effectively controlled by gate voltage. While trion intensity increases with gate voltages, exciton intensity decreases due to the conversion of excitons into trions with increasing electrons or holes. In both monolayer and bilayer cases, trion intensities show a minimum at negative voltages, denoted by $V_m$ (-2.5 V for bilayer and -1V for monolayer), indicating an initial existence of negative charges in ungated samples. Both PL and absorption spectra of monolayer samples show certain symmetric features with respect to $V_m$, where exciton feature is maximum. However, the corresponding spectral features of the bilayer sample show more complicated behavior. For almost all the bilayer $MoTe_2$ samples measured, there are two neutral excitons ($X_1$ and $X_2$) in both PL and absorption spectra that shift and convert with changing gate voltages. Similar two-exciton features in bilayer samples were also seen in $MoTe_2$ previously *(34)*. Trion absorption peak is only observable at very high gate voltages. We also noticed in Fig.1 (c) that the emission for hole-trions ($T^+$) for the bilayer sample is much stronger than that for electron-trions ($T^-$).

Figure 2 (a) shows PL spectra for the same bilayer sample as in Fig. 1 (c)&(d) as a function of excitation levels at 10 V. The PL spectra show predominantly two peaks centered at 1.137 eV and 1.124 eV, labeled as $X_2$ and $T^-$, which correspond to exciton and trion for bilayer $MoTe_2$, respectively, consistent with previous observations *(19,34-37)*. Figure 2 (b) shows a series of differential reflectance spectra with increasing excitation as processed using *equation (1)*. We see that there is a minimum bleaching of absorption near the two excitons ($X_1$ and $X_2$) within the relatively weak pump range. There is however a quite pronounced change of absorption features around 1.119 eV and eventually the absorption is over the background level (see more discussions in SM Section S2). Or optical gain occurs near 1.119 eV and increases with increasing pumping. A zoomed-in view of the gain spectra is shown in Fig. 2 (c), to clearly display the pump dependence and peak positions of the gain spectra. The optical gain reaches a maximum at a pump level of 30 μW. It should be noted that optical gain is different from the signal enhancement, which is a measure of absorption reduction, defined as absorption change with and without laser excitation (See SM Section S4 for more discussions).

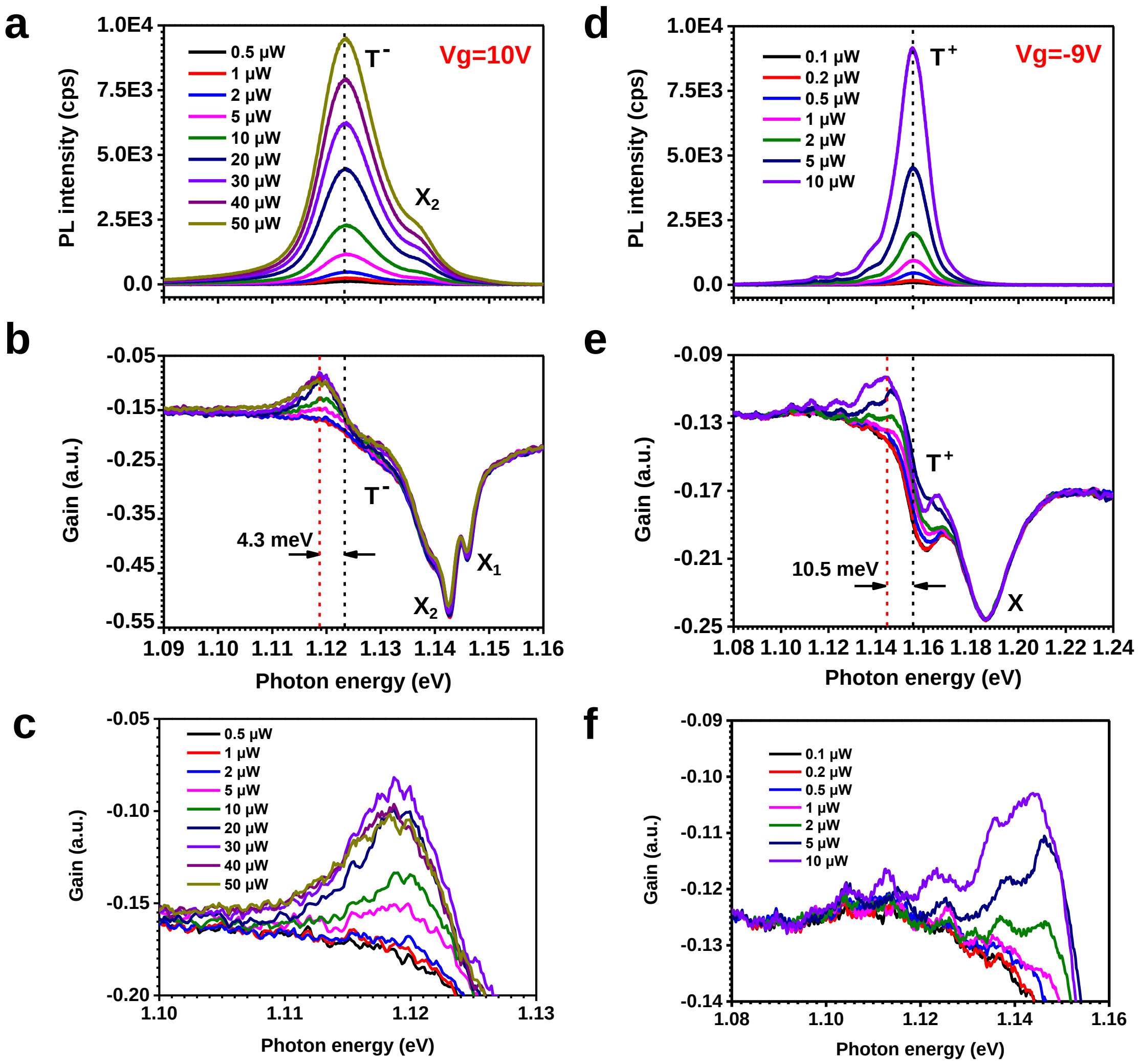

**Figure 2 | Photoluminescence and optical gain spectra.** Pump power dependent PL (**a**) and optical gain spectra (**b**) for a bilayer sample measured at a gate voltage of +10 V. The gain peak is ~ 4.3 meV below trion PL peak, as indicated by the two dashed lines. **c.** Zoomed-in view of figure **b** to more clearly display the pump dependence and peak positions. **d,e,f** Corresponding PL and gain spectra for the monolayer sample measured at a gate voltage of -9 V. The gain peak is ~ 10.5 meV below trion PL peak.

Figure 2 (d,e,f) show the results of similar measurement for the same monolayer sample as in Fig. 1 (e)&(f) at -9 V. The PL spectra are dominant by hole-trion emission ($T^+$), centered at 1.155 eV. Exciton emission is extremely weak and not visible (see also, Fig. 1(e)). Optical gain occurs near 1.145 eV. The appearance of gain with increasing pumping and the trend of change are very much the same as for the bilayer case. There is a general trend for both bi-layer and monolayer samples that can be described as follows: with the increase of pumping, the absorption features near trion energy starts to saturate (see

Fig. 2 (b)&(e)) and a peak emerges slightly below trion emission energy above the absorption background, indicating the appearance of an optical gain. For a given gate voltage, the gain peak saturates at a high enough pumping level (*e.g.*, 30 µW in Fig. 2(c)). However, the features of gain spectra become visibly more irregular for monolayer sample due to the existence of defect states below the trion peak, which are clearly visible in the PL spectra (see defect features in Fig. 2 (d) below 1.14 eV). We notice throughout this research that the gain spectrum becomes significantly noisy whenever defects become visible from the PL spectrum. This is especially the case for monolayer samples with significant defect emission. The noisy features in the gain spectrum (*e.g.*, in Fig. 2(f)) are mainly the results of error-amplification in the data-processing to obtain differential reflectance using *equation (1)*. Figure S7 in SM Section S5 shows the respective integrated PL intensities of excitons and trions as a function of pumping, showing nearly linear dependence. Such linear dependence excludes exciton-exciton scattering *(8)* or bi-exciton *(10,11)* as gain mechanism, since both feature quadratic dependence of PL on pumping. Within the range of pumping, no bi-exciton emission has been observed. Thus, the appearance of optical gain which is spectrally associated with trion state is likely originated from trions.

As typically the case, optical gain appears on the low energy side of PL peak and is ~ 4.3 meV and ~ 10.5 meV below trion peak for the bilayer and monolayer sample, respectively. This behavior and the relative spectral features strongly suggest that the optical gain originates from trions. The physical mechanism of trionic gain was first studied in doped ZnSe quantum wells *(12)*. According to this understanding, the trion system can be described by a "two-band" model: a ground state of a doped (*e.g.,* n-type) material which is the conduction band, $E_e$, filled with a given number of electrons (which could be pre-doped due to defects, gate generated or intentional doping) and a upper trion band, $E_T$, which has a much heavier effective mass ($m_T=2m_e+m_h$ for electron-trions), as illustrated in Fig. 3 (a). The trion formation is described more accurately by the "four-band" model as illustrated in Fig. 3 (b), which consists of three key steps: Excitons (bound electron-hole pairs) that are generated through optical pumping (step 1) quickly find their charged partners generated through gating (step 2) to form trions (step 3) by releasing a binding energy of $E_b^T$. With each pump generated exciton, one electron in the lower band is consumed to form a trion in the upper band. The net effect of a pumping photon (at energy $E_P$) is to decrease (increase) the population of lower band, $E_e$, (upper band, $E_T$) by one. Such a three-step process can lead to a population inversion between trion and electron bands and to achieve optical gain, as shown in step 3 of Fig. 3 (b). The upper limit of optical gain is reached when all pre-existing electrons find their exciton partners to form trions. The maximum gain would be limited by the total number of pre-doped electrons ( $n_D$ ). More precisely, the occupation of trion state (step 3) at a given pumping level ( $n_p$ ) is determined by the relative distribution of trions, electrons, and excitons (at $E_x$). Thus, the co-existence, mutual conversion, and the resulting steady-state distribution of free electrons, holes, and all excitonic complexes determine the population inversion and the amount of achievable optical gain.

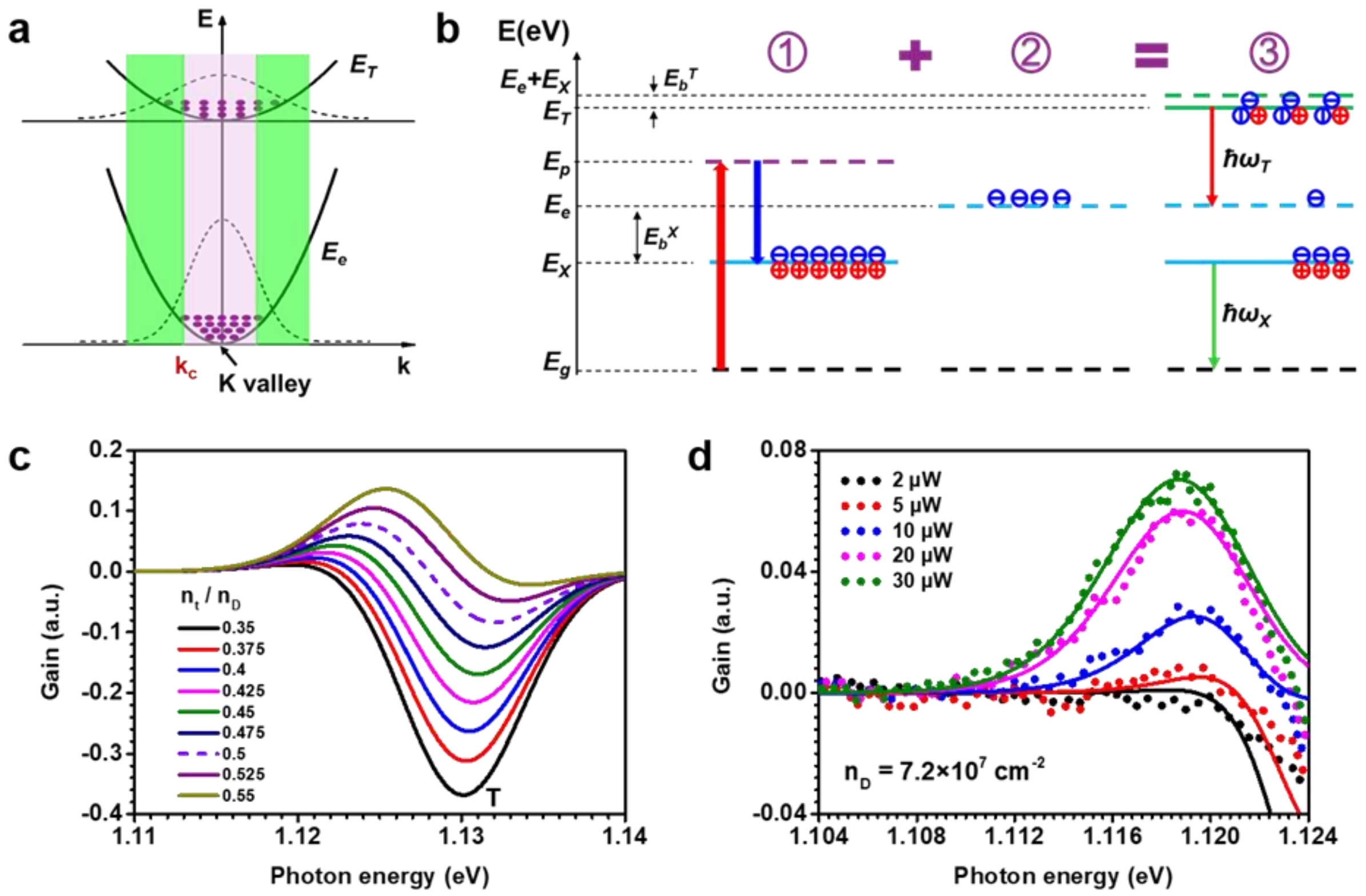

**Figure 3 | Physical mechanisms and theoretical model for trionic gain. a.** Parabolic bands (solid lines) and the electron distributions (dashed lines) in trion ($E_T$) and electron band ($E_e$). The pink inner band indicates a region around *K* valley where the absorption process is dominant, while in the outer green bands (separated at $k=k_c$) local population inversion can occur. **b.** Schematic of three key steps of trion formation through exciton generation ($E_X$) via optical pumping ($E_P$) from ground state ($E_g$) (step 1); the pre-existence of electrons ($E_e$) through gating or doping (step 2); and the possible population distribution among three states (trion, electron, and exciton) and the occurrence of population inversion (step 3). $E_b^T$ and $E_b^X$ denote binding energies for trion and exciton, respectively. **c.** Theoretical absorption and gain spectra at different values of ratio $n_t/n_D$ ($n_t$: trion density; $n_D$: doping level) from the model (*equation (3)*); **d.** Fitting result (solid lines) of the measured gain spectra (dot lines) from Fig. 2 (c) by the "four-band" model using *equation (3)*.

Interestingly, to achieve positive optical gain, there is no need for the total number of carriers in trion band to exceed the number in conduction band, or so-called global population inversion. As seen in Fig. 3 (a), due to the much heavier effective mass in trion state, the same number of carriers in trion bands occupies much larger range in *k*-space, leading to the existence of a cross-over *k* point, $k_c$, such that there is a local population inversion for $K-k_c > k > K+k_c$. This leads to a situation where the optical gain provided in the green bands could exceed the absorption in the pink band before the global inversion is reached, depending on pumping and linewidth broadening for a fixed level of dope density. Such occurrence of optical gain through local population inversion without global inversion was first observed in quantum cascade lasers *(38)*. But here the effect is more pronounced due to the much larger difference of effective masses. More quantitatively, optical gain is modeled by

$$G(\omega) \propto \sum_k |\mu_k|^2 L(k,\omega)(f_t - f_e) \quad (3)$$

where $f_t$ and $f_e$ stand for the fermi distributions of electrons in trion band and conduction band, respectively. $|\mu_k|^2$ is the optical dipole matrix element and $L(k,\omega)$ is a lineshape function (see SM Section S6 for details). The calculated gain spectra for a sequence of increasing ratio, $n_t / n_D$, are shown in Fig. 3 (c), where we see that the optical gain appears on the slightly red-side of trion peak energy, in consistence with experimental results. For a fixed electron doping density $n_D$, the critical condition for population inversion is $n_t = n_e = 0.5n_D$ for a typical two-band system. Since $n_D$ can be of any low value, in principle optical gain can occur at extremely low level of carrier density. This is in strong contrast to the occurrence of optical gain based on electron-hole plasma which requires very high transparency density. In addition, due to the possibility of local *k*-space population inversion before global inversion, optical gain occurs at a trion density slightly smaller than $0.5n_D$, as seen in Fig. 3 (c).

To understand our experimental measurement of gain spectrum more quantitatively in terms of the "four-band" model and the gain spectra calculated from *equation (3)*, we need to determine the stead-state distributions of electrons, excitons and trions using the well-known mass-action law *(3,39)*. For electron-trions formed by $X + e \leftrightarrow T^-$, we have $n_x n_e = n_t K(T)$, where $K(T)$ is a temperature-dependent equilibrium constant (see SM Section S7). From charge conservation, the trion density $n_t$ can be calculated as a function of $n_p$ and $n_D$. The pumping density, $n_p$, is determined by measured absorption coefficient at the pumping wavelength. The doping density, $n_D$, is determined by best-fitting experimentally measured spectra using *equation (3)* and calculated trion density (see SM Section S7). The experimental spectra agree with the modeled ones quite nicely with only a single $n_D$ (see in Fig. 3 (d)). It is important to emphasize that, the optical gain occurs below 5 µW pump power, corresponding to a pumping density of ~ $3.6\times10^7$ cm$^{-2}$. And the electron doping density obtained by best fitting was ~ $7.2\times10^7$ cm$^{-2}$, in good agreement with the "two-band" gain model discussed above in Fig. 3 (a), with all the populations determined by the "four-band" model in Fig. 3 (b) (step 3).

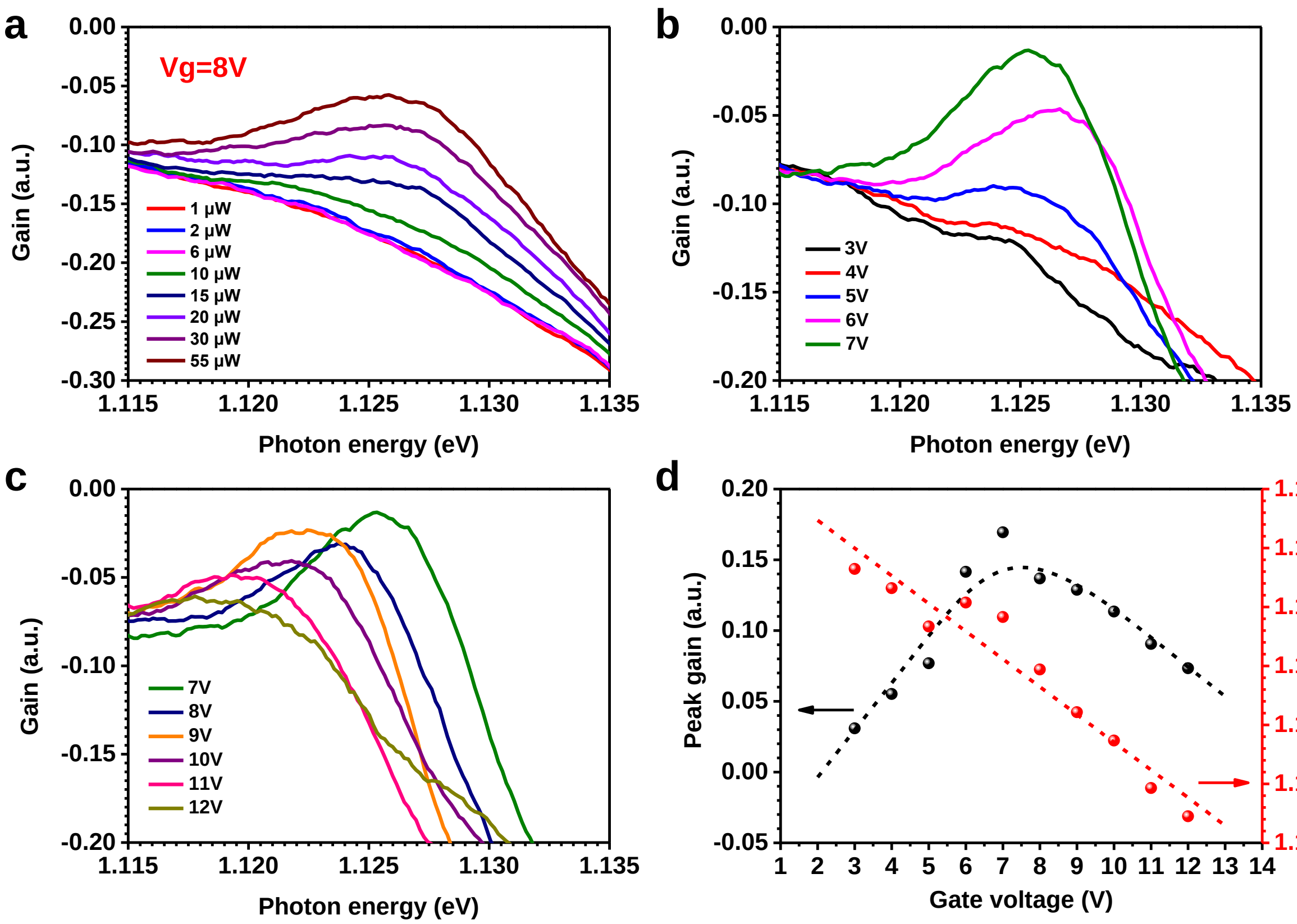

**Figure 4 | Optical gain of another bilayer sample. a.** Optical gain spectra at several pumping levels for another bilayer sample measured at 4K and a gate voltage of +8 V. **b, c** Optical gain spectra at several gate voltages at 40 μW pumping. The gain spectra are presented in two groups in figure **b** & **c** showing increasing (**b**) and decreasing (**c**) trend with gate voltages separately. **d.** Extracted peak gain values relative to the reflectance background and photon energies of the gain peaks as a function of gate voltages.

Figure 4 presents similar gain measurement for another bilayer $MoTe_2$ sample, where both pumping dependent gain spectra at a fixed gate voltage (a) and at varying gate voltages for a fixed pumping level (b,c,d) are shown. As shown in Fig. 4 (a), optical gain occurs at ~ 10 μW pumping ($n_p \sim 7.2 \times 10^7$ cm$^{-2}$). Clear pump dependent increase of optical gain is seen at relatively large gate voltage of 8 V, where high enough electron density ($n_D$) was produced by gating. With increasing pumping, increasing number of excitons are generated that find their partner electrons to form negative trions. This translates into more population in trion band, leading to a monotonous increase of optical gain. The gain evolution for various gate voltages at a fixed pumping level of 40 μW in Fig. 4 (b & c) are especially interesting. Optical gain initially increases with gate voltage from 3V to 7V. The 40 μW excitation level produced enough excitons for gate-generated electrons to form trions up to 7V. Thus with each increase of gate voltage, there is an increase of population in trion band, leading to an increase of optical gain in Fig. 4 (b). The gain starts to decrease when the gate voltage is further increased beyond 8V, as shown in Fig. 4 (c). This is because too many electrons are produced at higher gate voltages than the number of excitons produced at 40 μW. Trion population in the upper band is capped by the given total exciton numbers generated by pumping, while the electron population in lower band keeps increasing with the increased gate voltage beyond 8 V. Such decrease of population difference (compare to Fig. 3(b), step 3) due to the increased lower band population leads to continuous decrease of optical gain as gate voltage is further increased. This feature is more clearly displayed in Fig. 4 (d), where the extracted peak gain values relative to the reflectance background are plotted as increasing gate voltages, together with the photon energies of gain peaks. The trionic gain model and physical picture presented in Fig. 3 can explain the data here quite satisfactorily. More results at different gate voltages or for other devices are shown in Fig. S9 in SM Section S8 to provide more evidences of optical gain.

In summary, we have conducted systematic PL and reflectance spectroscopy experiments on electrically-gated $MoTe_2$ devices to investigate possible optical gain at low excitation levels. Our results on both mono- and bi-layer samples show clear existence of optical gain at such low carrier density levels, orders of magnitude below MD. After examining the signatures of optical gain and their relationship with excitons and trions, we could clearly identify the origin of optical gain as being trionic in nature. The physical model and understanding can explain our experimental results quite well. Our results have clearly established the existence and trionic origin of optical gain measured in 2D $MoTe_2$. We emphasize that the optical gain observed well below MD is of special interest in 2D materials. The main results and key conclusions of this paper are of great importance on several levels: First, while existence of optical gain and stimulated emission associated with excitonic complexes in conventional semiconductors have been occasionally studied and are of great importance both in basic condensed matter physics and in possible device applications, similar study has not been carried out as of present for 2D materials. Our results contribute significantly to the understanding of the nature of co-existence and mutual conversions of various excitonic complexes well below MD. Second, the existence of optical gain at extremely low-density levels is of tremendous importance for fabricating nanolasers with very low threshold and power consumption. As we pointed out above, the existence of optical gain does not depend on the total number of injected carriers, but on the population difference between trion band and electron band. Thus, the trion gain can in principle exist at any arbitrarily low level of carrier densities. This possibility opens a wide vista of making extremely low-threshold lasers down to single charge level *(13)*.

Importantly, there seems to be more intricate link between the existence of optical gain and MT. The recent study *(7)* revealed even more close relationship between "transparency" density ($N_t$) and MD ($N_m$), especially in low dimensional material systems. In most of conventional III-V or II-VI semiconductors, no optical gain typically exists in the low density regime, until a critical or "transparency" density is reached and optical gain appears subsequently when the carrier density exceeds $N_t$. Due to the small exciton binding energy in conventional semiconductors, we typically have $N_t > N_m$. Almost all the available lasers operate on this type of optical gain related to the degenerate EHP. $N_t$ is typically on the order of $10^{11}$ to $10^{12}$ $cm^{-2}$ depending on temperature, requiring high levels of excitation or energy input. Thus, it would be highly desirable to search for possible existence of optical gain at low density regimes, well below MD, *e.g.* $N<<N_m<N_t$. Such gain mechanisms would be extremely important for lower power, energy efficient device applications, given the current research interests of femto- or atto-joule optoelectronics *(40).* Excitonic related gain mechanisms for $N<N_m$ have only been demonstrated in rare cases for conventional semiconductors, including exciton-exciton scatterings *(8)*, localized excitons *(9)*, localized bi-excitons *(10)*, (free) bi-excitons *(11)*, trions *(12,13)*, and exciton-polariton scatterings *(14)*. This is why such lasers have not been widely used or manufactured. It is expected that all these rich varieties of gain mechanisms are now observable in 2D materials at room temperature and could find practical use in 2D materials-based lasers. At the same time, it is hoped that such study could reveal more profound links between various gain mechanisms and the co-existence or mutual conversions of various excitonic complexes in the vast range of low-to-intermediate-density regimes below MD.

As optical gain materials, 2D materials are particularly appealing for several reasons: first, the atomic-level thickness provides the thinnest optical gain materials to achieve potentially the smallest-volume low power lasers, since total power consumption of a nanolaser is determined by total volume of the gain material *(40)*; Second, the ultimate flexibility of 2D materials allows integration with strongly mismatched materials such as Silicon *(25)* for integrated photonics, without suffering from mechanical or thermal damages as in the case of III-V/Si integration. Such heterogeneous Si-2D material integration could be potentially superior to traditional III-V semiconductors that are the prevailing choices currently for silicon photonics, but faces many challenges due to their mechanical rigidity and small tolerance to lattice mismatches; And finally, the possibility of staggering different 2D materials to form artificial hetero-structures “by design” could be extremely important for eventually achieving 2D materials-based nanolasers under electrical injection, which represents a remaining challenge.

**Acknowledgments**
This research is supported by National Natural Science Foundation of China (Grant No. 20171302486), Beijing Innovation Center for Future Chips, Beijing National Center for Information Science and Technology, the 985 University Program, and Tsinghua University Initiative Scientific Research Program (No. 20141081296). The authors thank Prof. Yidong Huang's group for the usage of their fabrication equipment.

**Author contributions**
C.Z.N initiated the research and supervised the overall project. H.S. designed the reflectance spectroscopy experiments. H.S., Z.W., and Q.Y.Z. carried out the optical experiments. J.B.F, J.X.Z., Y.Z.L. participated in initial electrical device fabrication and the late devices were fabricated by Z.W. and Q.Y.Z. H.S., Z.W., Q.Y.Z., and C.Z.N. performed the data analysis. Z.W. and H.S. carried out the theoretical gain modeling. C.Z.N. proposed experimental search for the trion gain and the physical interpretation of gain mechanism. H.S. and C.Z.N. wrote the manuscript. All the authors participated in the discussions.

**Competing financial interests**
The authors declare no competing financial interests.

**Data availability**
The data that support the findings of this study are available from the corresponding author upon request.